\documentstyle[aps,manuscript]{revtex}

\begin{document}
\draft
\title{The Evaluation of the M{\o}ller Energy Complex in Difference 
Coordinate Representations}

\author{I-Ching Yang\footnote{E-mail:icyang@nttu.edu.tw} 
and Chia-Hsiu Tsai}

\address{Department of Natural Science Education, National Taitung University,\\ 
Taitung, Taiwan 950, Republic of China}

\maketitle

\begin{abstract}
The M{\o}ller energy complex of Schwarzschild black hole solution in several coodinates are evaluated.  Our results show that the M{\o}ller energy complex is independent of not only the purely spatial transformation, but also the shift of time coordinate.  So, we could conclude that a shift of time coordinates will not change the energy which is obtained by using the definition of M{\o}ller energy complex.
\end{abstract}

\vspace{2cm}
\noindent
\pacs{PACS No.:}
\newpage

To representing the physical system, we will choose suitable coordinates, like as the Cartesian coordinates, spherical coordinates or isotropic coordinates etc., according to given particular conditions.  Usually, these conditions are the properties of symmetry in system.  In general relativity, we could write down also the black hole solutions of various gravitational theories with diversiform coordinates for given reasons.  However, these reasons are not only the symmetry of space-time, but also to represent the geometrical structure of space-time.  Therefore, there are several kinds of coordinates are considered to express the black hole solutions in relativity.  Recently, many investigations~\cite{1} found out the energy distributions of black hole solutions in a few coordinates. While, for a satisfying definition of energy in curved space-time, there are various energy-momentum complexes, which include those of Einstein~\cite{2}, Tolman~\cite{3}, Papapetrou~\cite{4}, Bergmann~\cite{5}, Landau and Lifshitz~\cite{6}, and Weinberg~\cite{7}, be studied.  But they might depend on the choice of the spatial coordinates.  To introduce an energy-momentum complex that is independent of a purely spatial transformation, M{\o}ller constructed the following expression~\cite{8} 
\begin{equation}
\Theta^{\mu}_{\nu} = \frac{1}{8\pi} \frac{\partial \chi^{\mu\sigma}_{\nu}}
{\partial x^{\sigma}} ,
\end{equation}
where 
\begin{equation}
\chi^{\mu\sigma}_{\nu} = \sqrt{-g} \left( \frac{\partial g_{\nu\alpha}}
{\partial x^{\beta}} - \frac{\partial g_{\nu\beta}}{\partial x^{\alpha}}
\right) g^{\mu\beta} g^{\sigma\alpha} 
\end{equation}
is the M{\o}ller's superpotential, a quantity antisymmetric in the indices $\mu, \sigma$.  It is an interesting problem to see if the M{\o}ller energy complex enjoys the property of coordinate independence under a spatial transformation only.  On the other hand, by the evaluation of the energy of black hole solution with difference coordinates, we could study that what kinds of energy distributions are obtained while the several coordinate representations are used.  In this article, we evaluate the M{\o}ller energy complex of the black hole solution in several coordinate representations, and analyze these results.

Here we use the simplest solution of general relativity, the Schwarzschild black hole solution 
\begin{equation}
ds^2 = (1-\frac{2M}{r}) dt^2 - (1-\frac{2M}{r})^{-1} dr^2 - r^2 d\theta^2
- r^2 \sin^2 \theta d\phi^2 , 
\end{equation}
in our examination.  The Schwarzschild black hole solution is stationary and spherical symmetrical for vacuum field equations.  First, we consider the representation in Einstein-Rosen coordinates~\cite{9} 
given by the transformation rule:
\begin{equation}
(t, r, \theta, \phi)\longrightarrow (t, u=\sqrt{r-2M}, \theta, \phi) .
\end{equation}
Then, the line element is
\begin{equation}
ds^2 = \frac{u^2}{u^2+2M} dt^2 - 4(u^2+2M) du^2 - (u^2+2M)^2 (d\theta^2 +
\sin^2 \theta d\phi^2) . 
\end{equation}
The nonvanishing component of M{\o}ller's superpotential is 
\begin{equation}
\chi^{01}_0 = 2M \sin \theta .
\end{equation}
Next, we consider the representation in the Eddington-Finkelstein coordinates~\cite{10} as 
\begin{equation}
\bar{t} = t + 2M \ln |r-2M| ,
\end{equation}
and the line element can be written as
\begin{equation}
ds^2 = (1-\frac{2M}{r}) d\bar{t}^2 - \frac{4M}{r} d\bar{t} dr - \frac{r+2M}{r} 
dr^2 - r^2 d\theta^2 -r^2 \sin^2 \theta d\phi^2  .
\end{equation}
So, the nonvanishing component of M{\o}ller's superpotential is
\begin{equation}
\chi^{01}_0 = 2M \sin \theta .
\end{equation}
However, if the definition~\cite{10} 
\begin{equation}
(t, r, \theta, \phi)\longrightarrow 
(\bar{t} = t - 2M \ln |r-2M|, r, \theta, \phi) ,
\end{equation}
is chosen, the line element becomes
\begin{equation}
ds^2 = (1-\frac{2M}{r}) d\bar{t}^2 + \frac{4M}{r} d\bar{t} dr - \frac{r+2M}{r} 
dr^2 - r^2 d\theta^2 -r^2 \sin^2 \theta d\phi^2  .
\end{equation}
But, the nonvanishing component of M{\o}ller's superpotential is also
\begin{equation}
\chi^{01}_0 = 2M \sin \theta .
\end{equation}

Afterwards, we use a null coordinate~\cite{11} defined by
\begin{equation}
\tilde{v} = t + r + 2M \ln\left| \frac{r}{2M} -1\right| .
\end{equation}
In this representation, the line element is
\begin{equation}
ds^2 = (1-\frac{2M}{r})d\tilde{v}^2 - 2 d\tilde{v} dr - r^2 d\theta^2
- r^2 \sin^2 \theta d\phi^2  ,
\end{equation}
and nonvanishing M{\o}ller's superpotential is.
\begin{equation}
\chi^{01}_0 = 2M \sin \theta
\end{equation}
Fourth, we introduce the representation of the Painlev\'e-Gullstrand
coordinates~\cite{11} as follows
\begin{equation}
T = t +4M \left( \sqrt{\frac{r}{2M}} + \frac{1}{2} \ln \left| 
\frac{\sqrt{r/2M}-1}{\sqrt{r/2M}+1} \right| \right)  ,
\end{equation}
and the line element is given as 
\begin{equation}
ds^2 = (1-\frac{2M}{r}) dT^2 - 2\sqrt{\frac{2M}{r}} dT dr - dr^2 - r^2 d\theta^2 - r^2 \sin^2 \theta d\phi^2  .
\end{equation}
Again, the nonvanishing component of M{\o}ller's superpotential is
\begin{equation}
\chi^{01}_0 = 2M \sin \theta  .
\end{equation}
Finally, the Kruskal-Szekeres coordinates~\cite{11}, which is defined as 
\begin{eqnarray}
v & = & e^{r/4M} \cosh(t/4M)\sqrt{r/2M-1}  , \nonumber \\
u & = & e^{r/4M} \sinh(t/4M)\sqrt{r/2M-1}  .
\end{eqnarray}
for $ r> 2M $, is used to represent the Schwarzschild space-time out of
horizon. In Kruskal-Szekeres coordinates, the variable $v$ is timelike 
and $u$ is spacelike. Thus, the line element is reduced to the formula as
\begin{equation}
ds^2 = \frac{32M^3}{r} e^{-r/2M} (dv^2 - du^2) -r^2 d\theta^2 
- r^2 \sin^2 \theta d\phi^2  ,
\end{equation}
and the nonvanishing component of M{\o}ller's superpotential is
\begin{equation}
\chi^{01}_0 =  32 M^3 u e^{-r/2M} (1+\frac{r}{2M}) \sin \theta .
\end{equation}

The energy-momentum complex is evaluated by according to the definition of M{\o}ller
\begin{equation}
P_{\mu} = \frac{1}{8\pi} \int \frac{\partial \chi^{oi}_{\mu}}{\partial x^i} d^3 x.
\end{equation}
and the energy component is obtained by using the Gauss theorem 
\begin{equation}
E (r) = P_0 = \frac{1}{8\pi} \int \frac{\partial \chi_0^{0i}}
{\partial x^i} dr d \theta d \varphi  . 
\end{equation}
in which the Latin index also takes values from 1 to 3.  For the nonvanishing component of M{\o}ller superpotential in Eq.(6), (9), (12), (15), and (18), the M{\o}ller energy complex is 
\begin{equation}
E = M .
\end{equation} 
In Eq.(4), it is a purely spatial transformation, and its result correspond 
with the behavior of M{\o}ller energy-momentum pseudotensor.  The 
transformations of time coordinates of Eq.(7), (10), (13), and (16) is 
that new time coordinates are the original time coordinate add to arbitrary functions of $r$, as $f(r)$.  However, the M{\o}ller energy complex according to 
Eq.(19) is
\begin{equation}
E = 16 M^3 u e^{-r/2M} (1+\frac{r}{2M}) ,
\end{equation}
and this result is not a constant for time coordinate $u$.  In this case, 
we might say that the M{\o}ller energy complex is not independent of the case 
of Eq.(19).

According to our calculations, we could obtain that the M{\o}ller energy complex is independent of not only the purely spatial transformation, but also the shift of time coordinates, like as $T = t+f(r)$.  Following the calculation, we would try to give a more accurate proof which points out the M{\o}ller energy complex will not change under the shift of time coordinate. Let us to consider a static spherical coordinate system in which the line element has the form
\begin{equation}
ds^2 = A(r) dt^2 - B(r) dr^2 - r^2 d \theta^2 - r^2 \sin^2 \theta d \varphi^2  .
\end{equation}
The covariant metric is 
\begin{eqnarray}
g_{\mu\nu} = \left(  
\begin{array}{cccc}
A & 0 & 0 & 0 \\
0 & -B & 0 & 0 \\
0 & 0 & -r^2  & 0 \\
0 & 0 & 0 & -r^2 \sin^2 \theta 
\end{array}  \right)  .
\end{eqnarray}
After calculating, we get the nonvanishing component of M{\o}ller's superpotential
\begin{equation}
\chi_0^{01} = \frac{1}{\sqrt{AB}}\frac{\partial A}{\partial r} r^2 \sin \theta .
\end{equation}
On the other hand, we introduce a new time coordinate 
\begin{equation}
T = t + f(r)  ,
\end{equation}
which add a arbitrary shift $f(r)$ to original time coordinate.  Differentiating Eq.(30), we get
\begin{equation}
dT = dt+ f' dr ,
\end{equation}
where $f' \equiv  df/dr$, and substituting for $dt$ in Eq.(26). Then we get new line element as
\begin{equation}
ds^2 = A(r)dT^2 - 2A(r) f' dTdr - \left[ B(r) - A(r) f'^2 \right] dr^2
-r^2 d \theta^2 - r^2 \sin^2 \theta d \varphi^2  .
\end{equation}
At the same time, the covariant metric becomes
\begin{eqnarray}
g_{\mu\nu} = \left(  
\begin{array}{cccc}
A & -A f' & 0 & 0 \\
-A f' & -B+A f'^2 & 0 & 0 \\
0 & 0 & -r^2  & 0 \\
0 & 0 & 0 & -r^2 \sin^2 \theta 
\end{array}  \right)  ,
\end{eqnarray}
and its contravariant form becomes
\begin{eqnarray}
g^{\mu\nu} = \left(  
\begin{array}{cccc}
A^{-1}- f'^2 /B & -f' /B & 0 & 0 \\
-f' /B & -1/B & 0 & 0 \\
0 & 0 & -r^{-2}  & 0 \\
0 & 0 & 0 & -r^{-2} \sin^{-2} \theta 
\end{array}  \right)  .
\end{eqnarray}
The nonvanishing component of M{\o}ller's superpotential we obtained is 
\begin{equation}
\chi_0^{01} = \frac{1}{\sqrt{AB}}\frac{\partial A}{\partial r} r^2 \sin \theta  .
\end{equation}
These two different time coordinates in which a shift relation between give the same M{\o}ller's superpotential.  So, we could conclude that a shift of time coordinates will not change the energy which is obtained by using the definition of M{\o}ller energy complex.

\begin{center}
{\bf Acknowledgments}
\end{center}
This research was supported by the National Science Council of the Republic of China under contract number NSC 93-2112-M-143-001.

\end{document}